\documentclass[aps,prb,preprint,superscriptaddress,nobibnotes,amsmath,amssymb,amsfonts]{revtex4-1}

\usepackage{graphicx}
\usepackage{dcolumn}
\usepackage{bm}
\usepackage{multirow}
\usepackage{float}
\newcolumntype{C}[1]{>{\centering\arraybackslash}p{#1}}

\begin{document}

\title{Strain-driven superplasticity and modulation of electronic \\properties of ultrathin tin (II) oxide: A first-principles study}

\author{Devesh R. Kripalani}
\affiliation{School of Mechanical and Aerospace Engineering, Nanyang Technological University, Singapore 639798, Singapore}
\affiliation{Infineon Technologies Asia Pacific Pte Ltd, Singapore 349282, Singapore}

\author{Ping-Ping Sun}
\affiliation{School of Mechanical and Aerospace Engineering, Nanyang Technological University, Singapore 639798, Singapore}

\author{Pamela Lin}
\affiliation{Infineon Technologies Asia Pacific Pte Ltd, Singapore 349282, Singapore}

\author{Ming Xue}
\affiliation{Infineon Technologies Asia Pacific Pte Ltd, Singapore 349282, Singapore}

\author{Kun Zhou}
\email[]{kzhou@ntu.edu.sg}
\affiliation{School of Mechanical and Aerospace Engineering, Nanyang Technological University, Singapore 639798, Singapore}


\begin{abstract}
2D-layered tin (II) oxide (SnO) has recently emerged as a promising bipolar channel material for thin-film transistors and complementary metal-oxide-semiconductor devices. In this work, we present a first-principles investigation of the mechanical properties of ultrathin SnO, as well as the electronic implications of tensile strain ($\epsilon$) under both uniaxial and biaxial conditions. Bulk-to-monolayer transition is found to significantly lower the Young's and shear moduli of SnO, highlighting the importance of interlayer Sn-Sn bonds in preserving structural integrity. Unprecedentedly, few-layer SnO exhibits superplasticity under uniaxial deformation conditions, with a critical strain to failure of up to 74\% in the monolayer. Such superplastic behavior is ascribed to the formation of a tri-coordinated intermediate (referred to here as \textit{h}-SnO) beyond $\epsilon$ = 14\%, which resembles a partially-recovered orthorhombic phase with relatively large work function and wide indirect band gap. The broad structural range of tin (II) oxide under strongly anisotropic mechanical loading suggests intriguing possibilities for realizing novel hybrid nanostructures of SnO through strain engineering. The findings reported in this study reveal fundamental insights into the mechanical behavior and strain-driven electronic properties of tin (II) oxide, opening up exciting avenues for the development of SnO-based nanoelectronic devices with new, non-conventional functionalities.
\end{abstract}


\maketitle


\section{Introduction}
Of late, two-dimensional (2D) materials have piqued the curiosity of the scientific mind. From graphene\cite{NGgr04,ZTgr05} to transition metal dichalcogenides (TMDs),\cite{CPtmd14} this class of materials has received much acclaim for its role in next-generation nanoscale devices, with advanced functional applications in optoelectronics,\cite{MSopto16} molecular (gas) sensing,\cite{NCgassensor19} catalysis\cite{DNcatalysis16} and energy storage.\cite{ALenergy17} The physics of 2D materials, usually single- or few-atom thick, are highly non-trivial due to the significant effects of quantum confinement, enhanced structural bonding and strong charge coupling, giving rise to remarkable physical and chemical properties in the ultrathin limit. Many of these 2D candidates have been identified as semiconductors, with finite band gaps that can be manipulated under suitable structural modifications such as strain,\cite{KKstrain18} stacking registry,\cite{DZstack14} defects and functionalization.\cite{SFdefects17}

\begin{figure*}
\includegraphics[width=0.96\textwidth]{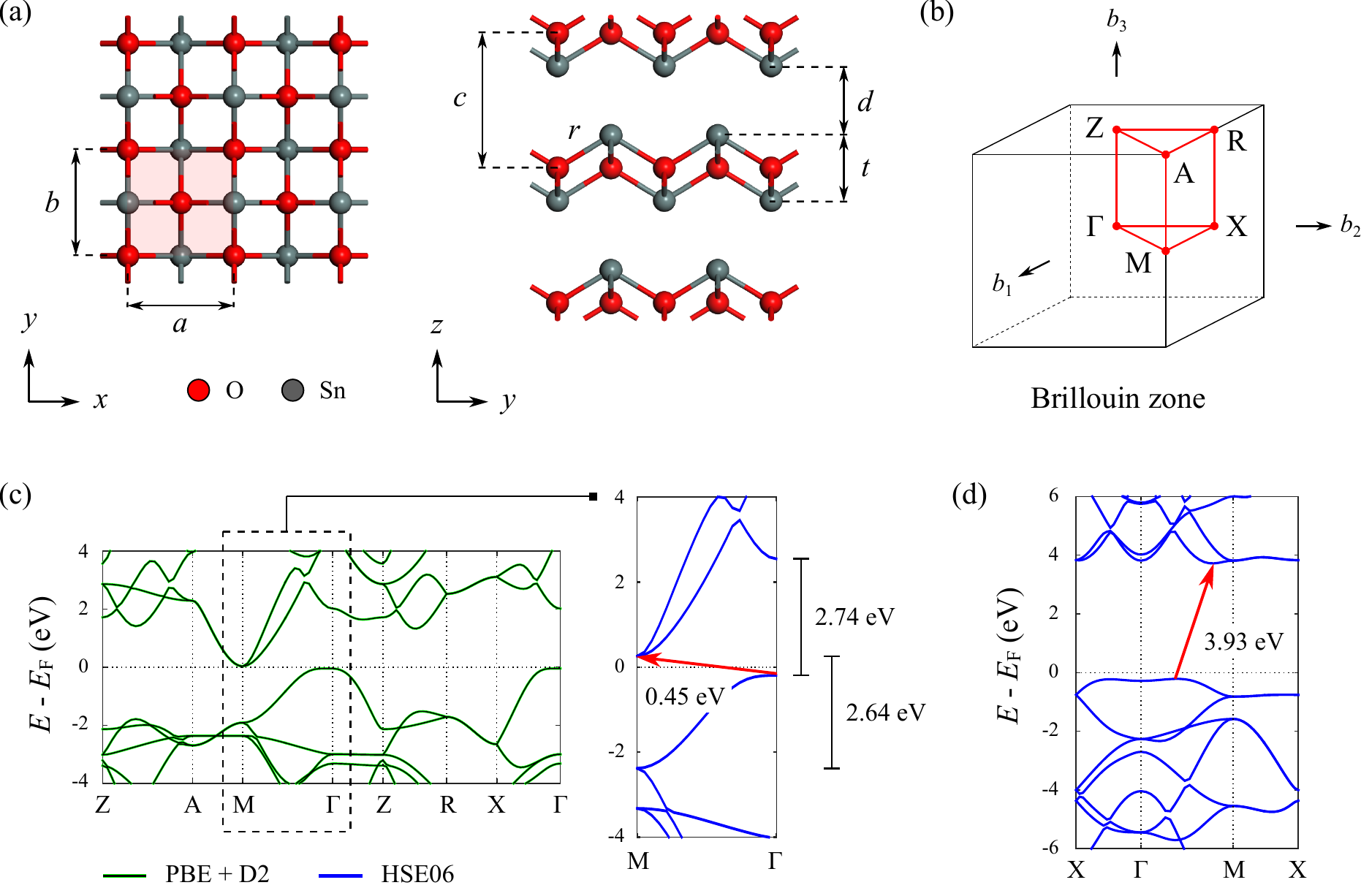}
\caption{(a) Crystal structure of tin (II) oxide (SnO), with the tetragonal unit cell (shaded pink) indicated in the \textit{xy}-plane. (b) First Brillouin zone and its corresponding high symmetry paths (red). Electronic band structures of (c) bulk and (d) monolayer SnO based on PBE+D2 (black-green) and HSE+D2 (blue) methods. \label{F1}}
\end{figure*}

Recently, 2D-layered tin (II) oxide (SnO) has emerged as a promising candidate for thin-film transistors (TFTs)\cite{OHtfta08,CNtftb13,KBtftc17} and complementary metal-oxide-semiconductor (CMOS) devices.\cite{NKcmosa11,YTcmosb12} SnO is a typical van der Waals layered compound with Sn-O-Sn atomic planes stacked along the [001] crystal orientation (see Fig. \ref{F1}(a)). Given its relatively narrow band gap in bulk ($\sim$0.7 eV), recent progress in isolating ultrathin films of SnO in a bid to open up the band gap has greatly amplified its technological impact in fields like photocatalysis and transparent conductivity. Many theoretical works have also contributed towards an atomistic understanding of SnO, for example in relation to the role of native defects,\cite{TOnatdef06} transition metal doping possibilities\cite{WLtmdoped18} and its surface adsorption characteristics.\cite{TGadschar17} Furthermore, unlike most 2D chalcogenides which are highly degradable in ambient conditions, metal oxide-based systems as such are generally regarded to be resistant to oxidation.\cite{GGoxide2d13,GMoxide2d19}

High-quality few-layer SnO has been successfully obtained experimentally via conventional thin-film fabrication methods such as atomic layer deposition\cite{KBtftc17} and electron beam evaporation.\cite{PLebeam16} In a study by Saji \textit{et al}., the growth of SnO has also been reportedly achieved with single-layer precision on sapphire and SiO$_2$ substrates using pulsed laser deposition.\cite{STpld16} Tin (II) oxide has been widely explored as a \textit{p}-type channel layer for high-performance TFTs, with field effect mobilities of as high as 1.9 cm$^2$V$^{-1}$s$^{-1}$.\cite{STpld16} Great strides have also been made to realize ambipolar behaviors in SnO for complementary-like operation,\cite{NKcmosa11,YTcmosb12} while complete carrier polarity switching (\textit{n}-type) has also been demonstrated through doping with antimony (Sb).\cite{HOdopesb11} Previous studies have shown that the electronic band structure of metal oxides tend to be sensitive to mechanical strain.\cite{TGoxbga08,JSoxbgb11,WGoxbgc14,ZLoxbgd14} In practical nanoscale devices, common sources of strain include lattice mismatch at contact interfaces and the integration with flexible pre-strained substrates as part of strain engineering applications. However, in the case of atomically thin SnO, the effects of strain on its structure and (opto)electronic properties are yet to be fully understood despite being an inevitable issue of real concern.

In this work, the mechanical properties of ultrathin SnO, as well as the electronic implications of tensile strain under both uniaxial and biaxial conditions, are investigated from first-principles calculations. Insights into the underlying deformation process, especially when equilibrium lattice symmetry is no longer preserved, can prove useful for achieving new, non-conventional functionalities of tin (II) oxide. Additionally, the influence of strain on its work function and band alignment is explored, offering fresh ground for facilitating efficient charge transport across SnO-based heterojunctions and interfaces with metals.

\section{Computational Method}
Spin-polarized first-principles calculations are performed within the framework of density functional theory (DFT)\cite{KSexcorr65,SSdft11} using the Vienna \textit{ab initio} simulation package (VASP).\cite{KFvasp96} The Perdew-Burke-Ernzerhof (PBE)\cite{PBE96} and hybrid Heyd-Scuseria-Ernzerhof (HSE06)\cite{HSE03,HSE06} exchange-correlation functionals are adopted in this work, along with the DFT-D2 method of Grimme\cite{DFTD206} to account for the van der Waals forces in layered tin (II) oxide. A kinetic energy cutoff of 500 eV is selected for the plane wave basis set. Bulk and few-layer structures are respectively sampled with 10 $\times$ 10 $\times$ 8 and 10 $\times$ 10 $\times$ 1 \textit{k}-point grids in the Brillouin zone using the Monkhorst-Pack method. The energy convergence criteria for electronic iterations is set at 10\textsuperscript{-6} eV and all structures are relaxed until the maximum force per atom is smaller than 0.01 eV/\AA.

To model 2D systems, periodic boundary conditions are applied in both in-plane (\textit{x} and \textit{y}) directions, whereas free boundary conditions are enforced in the normal (\textit{z}) direction by introducing a vacuum layer of $\sim$20 \AA\ to eliminate spurious interactions between neighboring slabs. By convention, strained conditions are prescribed according to the relation \textit{a}$_1$ = \textit{a}$_0$(1+$\epsilon$), where \textit{a}$_0$ and \textit{a}$_1$ are the lattice constants in equilibrium and at a strain level of $\epsilon$, respectively. In the case of applying uniaxial strain, the lattice constant in the transverse in-plane direction is fully relaxed to minimize the forces acting in this direction.

\section{Results and Discussion}
\subsection{Characterization of SnO in the elastic regime}

\begingroup
\squeezetable
\begin{table*}[htbp]
\caption{The relaxed geometric parameters and elastic constants of bulk and monolayer tin (II) oxide (SnO), as calculated in this work. \label{relaxed_params}}
\begin{ruledtabular}
{\renewcommand{\arraystretch}{1.5}
\begin{tabular}{ccccccc}
\multirow{2}{*}{System} & \multicolumn{2}{c}{\multirow{2}{*}{Geometric parameters}} & \multicolumn{4}{c}{Elastic constants} \\ \cline{4-7}
& & & \multicolumn{2}{c}{In-plane (\textit{xy})} & \multicolumn{2}{c}{Normal (\textit{z})} \\ \hline
Bulk & \textit{a} = \textit{b} = 3.85 \AA & \textit{c} = 4.87 \AA & \textit{E}$_{x}$ = 51 GPa & \textit{E}$_{y}$ = 51 GPa & \textit{E}$_{z}$ = 51 GPa & \\
SnO & \textit{t} = 2.34 \AA & \textit{d} = 2.52 \AA & \textit{G}$_{xy}$ = 87 GPa & & \textit{G}$_{xz}$ = 45 GPa & \textit{G}$_{yz}$ = 45 GPa \\
& \textit{r} = 2.25 \AA & & $\nu_{xy}$ = 0.72 & $\nu_{yx}$ = 0.72 & $\nu_{zx}$ = 0.13 & $\nu_{zy}$ = 0.13 \\
& & & $\nu_{xz}$ = 0.13 & $\nu_{yz}$ = 0.13 & & \\
& & & & & & \\
Monolayer & \textit{a} = \textit{b} = 3.83 \AA & & \textit{E}$_{x}$ = 33 GPa & \textit{E}$_{y}$ = 33 GPa & & \\
SnO & \textit{t} = 2.39 \AA & & \textit{G}$_{xy}$ = 79 GPa & & & \\
& \textit{r} = 2.26 \AA & & $\nu_{xy}$ = 0.81 & $\nu_{yx}$ = 0.81 & & \\
& & & $\nu_{xz}$ = 0.10 & $\nu_{yz}$ = 0.10 & &
\end{tabular}}
\end{ruledtabular}
\end{table*}
\endgroup

Tin (II) oxide adopts a tetragonal (litharge-type) crystal structure with space group \textit{P}4/\textit{nmm} (no. 129) in the bulk state, as shown in Fig. \ref{F1}(a). We report optimized lattice constants \textit{a} = \textit{b} = 3.85 \AA\ and \textit{c} = 4.87 \AA\ based on the PBE functional in conjunction with the DFT-D2 method (i.e. PBE+D2), which are in excellent agreement with experimental values (\textit{a} = \textit{b} = 3.80 \AA, \textit{c} = 4.84 \AA).\cite{PDsnoexpa80,GPsnoexpb06} Its internal stacking registry comprises of Sn-O-Sn monolayers in direct vertical alignment with one another along the [001] direction (interlayer separation \textit{d} = 2.52 \AA). Within each layer, the Sn-O bond length \textit{r} is found to be 2.25 \AA, while the core-to-core thickness \textit{t} is 2.34 \AA. The planar structure of SnO is noted to be insensitive to the number of layers; on thinning down into its monolayer, \textit{a} = \textit{b} = 3.83 \AA, \textit{r} = 2.26 \AA\ and \textit{t} = 2.39 \AA. A summary of the geometric parameters for both bulk and monolayer SnO, as calculated in this work, is given in Table \ref{relaxed_params}.

Next, we evaluate the effect of quantum size on the electronic band structure of SnO. Figure \ref{F1}(b) shows the first Brillouin zone corresponding to the tetragonal unit cell, along with paths (red) connecting the internal points of high symmetry. Benchmark calculations for the bulk system are presented in Fig. \ref{F1}(c) to assess the use of PBE+D2 (black-green) and HSE+D2 methods (blue). The PBE+D2 approach gives reasonably good predictions of the band dispersion and indirect nature of the band gap ($\sim$0.08 eV) in bulk SnO. However, since the PBE functional lacks a complete description for the self-interaction of electrons, hybrid functional methods are usually necessary for reproducing band gaps that are quantitatively consistent with experimental results. Refined calculations along M-$\Gamma$ with the HSE+D2 approach verify the indirect band gap in bulk SnO to be 0.45 eV, which occurs between the valence band maximum (VBM) at the $\Gamma$ point and conduction band minimum (CBM) at the M point. Furthermore, the optical (direct) band gap is estimated to be around 2.74 eV (2.64 eV) at the $\Gamma$ (M) point, enabling SnO to exhibit high transparency ($>$80\%) in the visible range.\cite{SMsnoopt93} Our findings here agree well with experimental works on SnO that have quoted the indirect and direct band gaps as $\sim$0.7 eV\cite{OHtfta08} and in the range of 2.5 - 3.3 eV,\cite{OHtfta08,SMsnoopt93,GRsnoopt84} respectively. We also see considerable band dispersion at the top-most valence band along $\Gamma$-Z (PBE+D2 result), an artefact of the presence of strong interlayer lone-pair (Sn-5\textit{s}) interactions.\cite{ZUsnolp15} The relatively large curvature of the VBM along Z, as compared to along X and M, suggests anisotropic \textit{p}-type conduction in bulk SnO with higher hole mobility in the interlayer rather than intralayer region. In the monolayer, the indirect band gap widens significantly to 3.93 eV with both the VBM and CBM located at distinct points along $\Gamma$-M, as shown in Fig. \ref{F1}(d).

Under small deformations, for which Hooke's Law is valid, the mechanical properties of tin (II) oxide can be characterized based on its elastic constants. The deformation response $\bm{\epsilon}$ of a given material subject to a state of stress $\bm{\sigma}$ is described by the generalized stress-strain relation $\bm{\epsilon}$ = \textbf{S}$\bm{\sigma}$ (or equivalently $\bm{\sigma}$ = \textbf{C}$\bm{\epsilon}$, where \textbf{C} = \textbf{S}$^{-1}$). Here, \textbf{S} and \textbf{C} denote the compliance and stiffness matrices, respectively. They contain the material-dependent elastic constants, namely the Young's modulus \textit{E}, Poisson's ratio $\nu$ and shear modulus \textit{G}. Details on our formulation for material characterization in the elastic regime are provided in Section I of the Supplemental Material (SM).\cite{SuppMater} Note that for 2D systems, plane stress conditions ($\sigma_{z}$ = $\tau_{yz}$ = $\tau_{xz}$ = 0) are imposed due to free relaxation of the surface in the normal (\textit{z}) direction. The stiffness matrices of bulk (\textbf{C}\textsuperscript{Bulk}) and monolayer (\textbf{C}\textsuperscript{ML}) SnO are presented in Eqs. (\ref{C6}) and (\ref{C3}), while their corresponding elastic constants are listed in Table \ref{relaxed_params}.
{\small
\begin{gather}\label{C6}
\textbf{C}\textsuperscript{Bulk} =
\begin{bmatrix}
118.99 & 89.24 & 27.10 & 0 & 0 & 0 \\
 & 118.99 & 27.10 & 0 & 0 & 0 \\
 & & 58.05 & 0 & 0 & 0 \\
 & & & 44.96 & 0 & 0 \\
\multicolumn{3}{c}{Sym.} & & 44.96 & 0 \\
 & & & & & 87.20
\end{bmatrix}
\end{gather}
}
{\small
\begin{gather}\label{C3}
\textbf{C}\textsuperscript{ML} =
\begin{bmatrix}
94.31 & 75.93 & 0 \\
 & 94.30 & 0 \\
Sym. & & 78.66
\end{bmatrix}
\end{gather}
}

Clearly, we see that SnO behaves as a transversely isotropic material in the surface (\textit{xy}) plane regardless of thickness. The bulk modulus \textit{K} under hydrostatic pressure is computed as 49 GPa, while the in-plane linear stiffness (\textit{B}$_x$ = \textit{B}$_y$ = 343 GPa) is found to be nearly five times higher than that along the normal direction (\textit{B}$_z$ = 69 GPa). These results are consistent with data obtained from prior experimental characterization of SnO by X-ray diffraction (\textit{K} = 35 - 53 GPa).\cite{GPsnoexpb06,ACsnobma92,MOsnobmb01} Interestingly, the transition from bulk to monolayer does not improve the in-plane mechanical properties of SnO as one would typically expect of 2D materials in the absence of interlayer slippage effects.\cite{TSelas2D18} The smaller Young's and shear moduli of the monolayer (see Table \ref{relaxed_params}) instead indicates a depreciation in its intrinsic resistance against elastic deformation, thereby highlighting the importance of interlayer Sn-Sn bonds in preserving the structural integrity of SnO. Akin to phosphorene (\textit{E} = 44 - 166 GPa)\cite{WPbpmech14} and antimonene (\textit{E} = 104 GPa),\cite{KKstrain18} ultrathin SnO exhibits a Young's modulus that is relatively low as compared to graphene (\textit{E} = 1 TPa)\cite{LWgrmech08} and MoS$_2$ (\textit{E} = 0.33 TPa).\cite{CPtmdmech12} Such low elastic rigidity may endow SnO thin films with high flexibility and wide compatibility across various substrates, opening up promising applications in practical strain-based electronics and nanoscale devices.

\begin{figure*}
\includegraphics[width=0.96\textwidth]{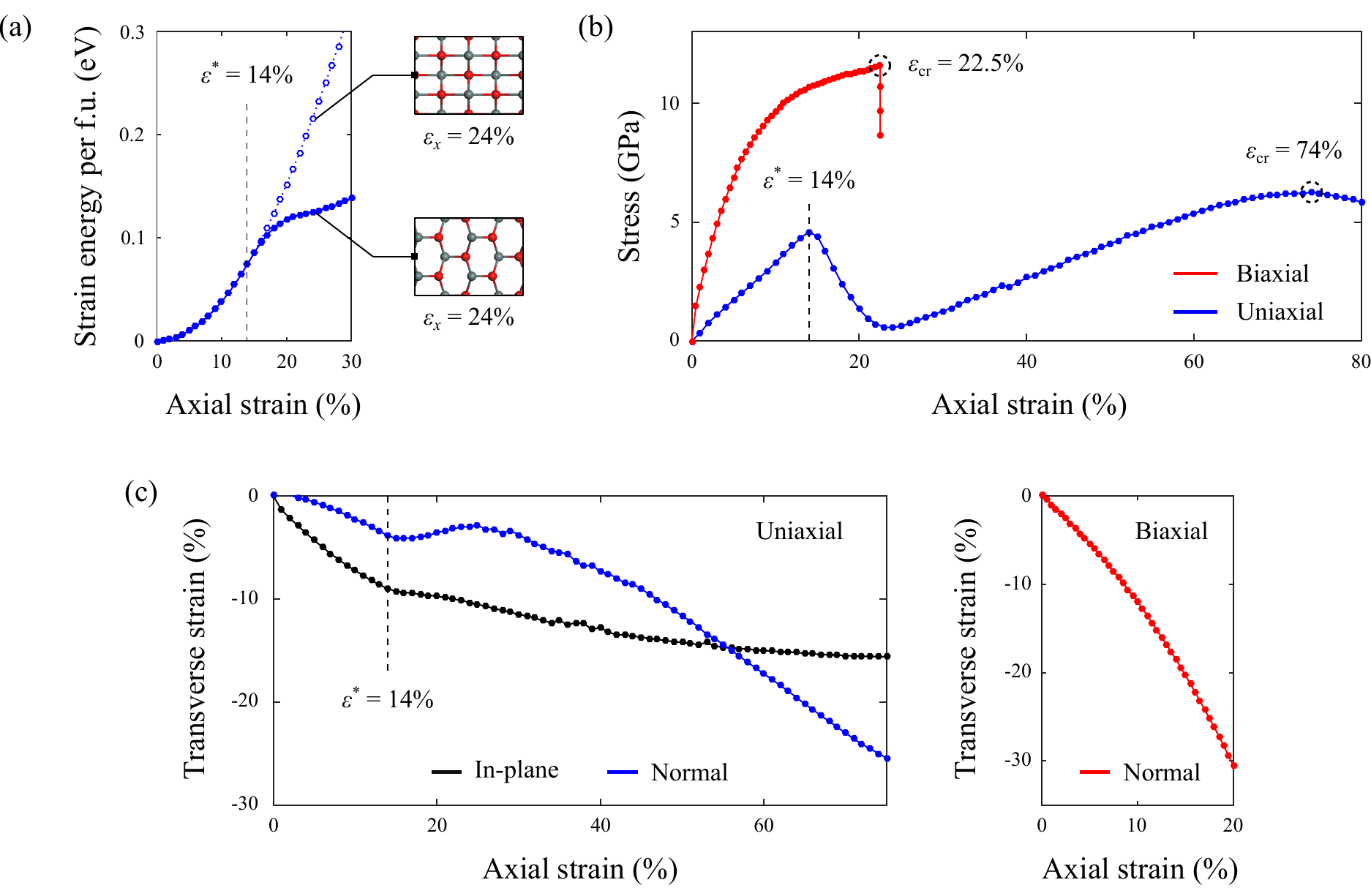}
\caption{(a) The evolution of strain energy per formula unit (f.u.) for monolayer SnO subject to uniaxial strain. Solid and dotted lines (blue) correspond to the cases with and without full internal structural optimization respectively, while snapshots of the lattice at $\epsilon_{x}$ = 24\% are given in the inset. (b) The stress-strain relationship, and (c) transverse relaxation response of monolayer SnO under applied biaxial and uniaxial strain conditions. The transition from low to high strain regimes in the uniaxial mode is denoted by dotted lines at $\epsilon$* = 14\%. \label{F2}}
\end{figure*}

\begin{figure*}
\includegraphics[width=0.96\textwidth]{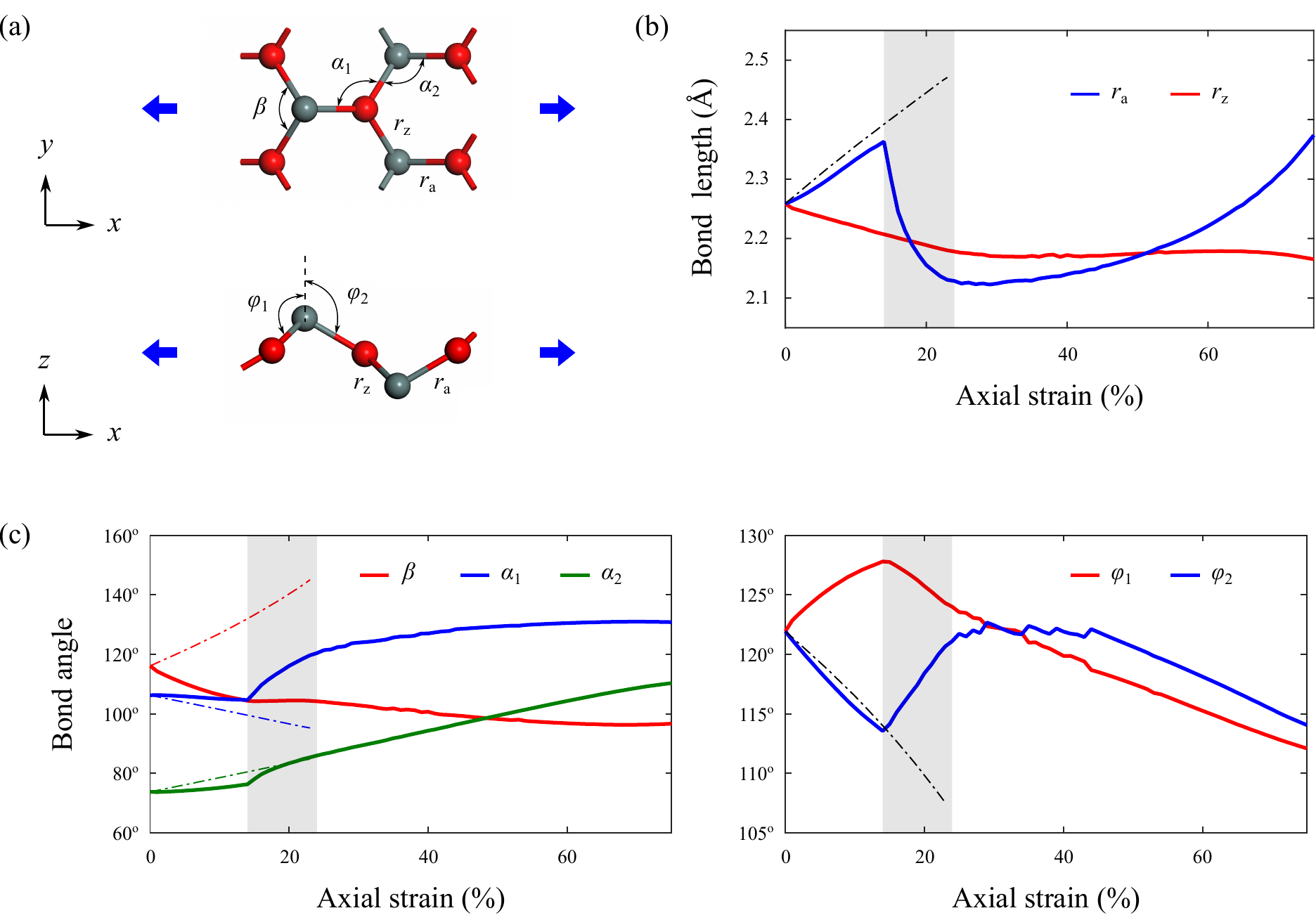}
\caption{(a) Representative orthorhombic structure of monolayer SnO under uniaxial tension along the \textit{x}-axis (blue arrows). The variation in (b) bond lengths, and (c) bond angles during deformation under applied uniaxial strain conditions. The grey-shaded region represents the formation window (14\%-24\%) of \textit{h}-SnO. Dash-dotted lines indicate the associated deformation response of the monolayer for the case of biaxial strain, where \textit{r}\textsubscript{a} = \textit{r}\textsubscript{z} and $\varphi_{1}$ = $\varphi_{2}$. \label{F3}}
\end{figure*}

\subsection{Ideal tensile deformation and superplasticity in monolayer SnO}
As the cornerstone of modern engineering at the nanoscale, the deliberate application of strain provides a useful means for achieving desired optical and electronic properties in low-dimensional materials. In this section, the effects of (i) biaxial ($\epsilon_{x}$ = $\epsilon_{y}$), and (ii) uniaxial ($\epsilon_{x}$) tensile strain on the structural deformation of monolayer SnO is examined. Note that due to the tetragonal symmetry of the lattice, the deformation response under uniaxial strain $\epsilon_{y}$ is identical to that of $\epsilon_{x}$ such that a complementary understanding can be intuitively drawn from (ii). The case of uniaxial strain presents a particularly interesting situation whereby the crystal structure of SnO transits into a lower-symmetry orthorhombic phase. It has previously been established that the orthorhombic form of SnO (also known as $\gamma$-SnO) can remain stable in its bulk state under high-pressure compression.\cite{ACsnobma92,LLsnohp07} However, its stability under tension, especially in the 2D limit (i.e. monolayer), continues to pose an open question of interest.

The evolution of strain energy per formula unit (f.u.) for monolayer SnO subject to uniaxial strain is shown in Fig. \ref{F2}(a). We find that the interatomic bonding configuration within the lattice is well-preserved in the low-strain regime of up to $\epsilon$* = 14\%. However, with increasing tension, Sn-O bonds oriented along the direction of strain start to weaken, leading to significant internal restructuring while minimizing the strain energy. As a result, a new honeycomb-like lattice (referred to here as \textit{h}-SnO) with low bond density begins to take shape at higher levels of strain beyond 14\%. In this structure, each Sn atom is no longer bonded to four, but to three adjacent O atoms, and vice versa (see inset of Fig. \ref{F2}(a) for the case of $\epsilon_{x}$ = 24\%). This configuration has its origin as a lesser-known polymorphic form of SnO, with equilibrium lattice constants \textit{a} = 5.01 \AA\ and \textit{b} = 3.36 \AA, as reported in a recent computational study by Guo \textit{et al.}\cite{GMoxide2d19} The energetic benefit, given by the difference between the hypothetical case without internal restructuring (dotted line) and the fully optimized structure (solid line), increases to around 0.1 eV per f.u. at $\epsilon_{x}$ = 24\%. For the \textit{h}-SnO phase, armchair and zigzag bonding patterns may also be accordingly identified along the direction of applied strain (\textit{x}) and along the transverse in-plane direction (\textit{y}), respectively.

The theoretical stress-strain relationship and transverse relaxation response of monolayer SnO under both biaxial and uniaxial strain conditions are presented in Fig. \ref{F2}(b)-(c). The tensile stress $\sigma$ is determined from the derivative of strain energy \textit{E}\textsubscript{s} with respect to strain according to Eq. (\ref{applstress}), where \textit{V} is the instantaneous volume of the system at a given strain level of $\epsilon$.
{\small
\begin{align}\label{applstress}
\sigma(\epsilon) = \frac{1}{\textit{V}(\epsilon)} \bigg(\frac{\partial\textit{E}\textsubscript{s}}{\partial\epsilon}\bigg)
\end{align}
}

We report critical failure strain (stress) values of 22.5\% ($\sim$11.6 GPa) and 74\% ($\sim$6.2 GPa) for the biaxial and uniaxial deformation mode, respectively. Most notably, we find that the critical strain for the uniaxial case is more than three times than that for the biaxial case. Such superplasticity under uniaxial conditions is unprecedented and may be ascribed to the formation (and subsequently, failure) of the tri-coordinated \textit{h}-SnO intermediate which we have disclosed earlier. Our results are supported by a similar set of calculations for the case of the bilayer (see Fig. S1 in the SM),\cite{SuppMater} where we note lower critical strain values (12.5\% and 68\% for the biaxial and uniaxial deformation mode, respectively) owing to structural instabilities arising from interlayer interactions. Nevertheless, it remains likely that the superplastic behaviour of SnO may be sustained well beyond the monolayer, with strong practical implications on the mechanical properties and performance of ultrathin SnO films in application. Here, it must be pointed out that, in the scope of the present study, the superplastic potential of SnO is interpreted simply in terms of its deformation behaviour as a pristine monocrystal, and does not account for other, possibly more dominant, processes known to occur in real polycrystalline samples. Mechanisms responsible for enhanced plasticity in both 3D\cite{VL3dsp94,ZT3dsp05,PC3dsp09,SB3dsp10} and 2D\cite{DJ2dsp07,CL2dsp14,ZL2dsp15} materials are generally understood to be thermally-activated and may include one or more of the following: grain boundary sliding, grain reorientation, dislocation-related activity, diffusional flow and dynamic recrystallization. Further experimental and theoretical investigation is therefore required to evaluate the role of such mechanisms in atomically thin tin (II) oxide, especially in light of its intrinsic capacity to accommodate encouragingly high uniaxial strains through a tetragonal-to-orthorhombic phase transformation.

For both biaxial and uniaxial cases, applied tensile strain leads to a corresponding contraction of the lattice in the transverse direction(s), in line with earlier findings on the positive Poisson's ratios of monolayer SnO (see Table \ref{relaxed_params}). Under low-to-moderate uniaxial strain, contraction along the transverse in-plane direction tends to dominate the relaxation response. However, this trend gradually diminishes with the onset of the \textit{h}-SnO phase beyond $\epsilon$* = 14\%, for which relaxation occurs more freely in the normal direction.

To gain further insight into the structural transformation and superplasticity of monolayer SnO under uniaxial strain, we investigate the evolution of its orthorhombic lattice during deformation, as defined by the geometric parameters in Fig. \ref{F3}(a). The variation in Sn-O bond lengths in the armchair (\textit{r}\textsubscript{a}) and zigzag (\textit{r}\textsubscript{z}) direction is given in Fig. \ref{F3}(b), while changes in the dihedral ($\alpha_{1}$, $\alpha_{2}$, $\beta$) and bond-to-normal angles ($\varphi_{1}$, $\varphi_{2}$) are plotted in Fig. \ref{F3}(c). Note that for the more straightforward case of biaxial strain (see dash-dotted lines), the tetragonal symmetry of the lattice is preserved such that \textit{r}\textsubscript{a} = \textit{r}\textsubscript{z} and $\varphi_{1}$ = $\varphi_{2}$ throughout the deformation process. Conversely, under uniaxial conditions, structural changes are strongly mediated by relaxation effects in the transverse in-plane and normal directions. The formation of \textit{h}-SnO is shown to take place over a broad window (14\% - 24\%), as evidenced by characteristic trend reversals in \textit{r}\textsubscript{a}, $\varphi_{1}$ and $\varphi_{2}$, which indicate the partial recovery of Sn-O bonds along the direction of applied strain. Apart from its contribution towards superplasticity, mechanically excited \textit{h}-SnO with optimal bond strength may even be obtained at $\sim$24\%, opening up intriguing possibilities for designing novel SnO-based nanostructures through strain engineering. In this regard, suitable means for promoting the stability of \textit{h}-SnO may be necessary, for example, through the careful selection of substrate/contact materials, surface functionalization and/or environment control.

\begin{figure*}
\includegraphics[width=0.96\textwidth]{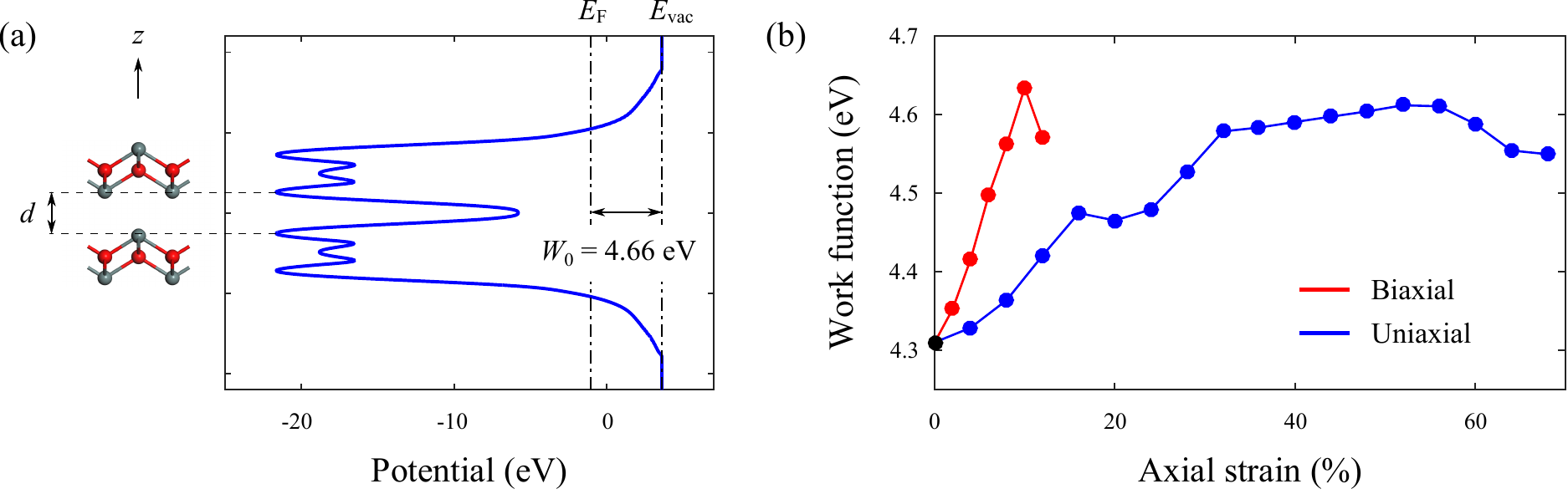}
\caption{(a) Planar-averaged potential of bilayer SnO in its equilibrium configuration, where \textit{E}\textsubscript{F} and \textit{E}\textsubscript{vac} are the energy values at the Fermi and vacuum levels, respectively (Level of theory: HSE+D2). (b) The variation of the work function under applied biaxial and uniaxial strain conditions (Level of theory: PBE+D2). \label{F4}}
\end{figure*}

\begin{figure*}
\includegraphics[width=0.96\textwidth]{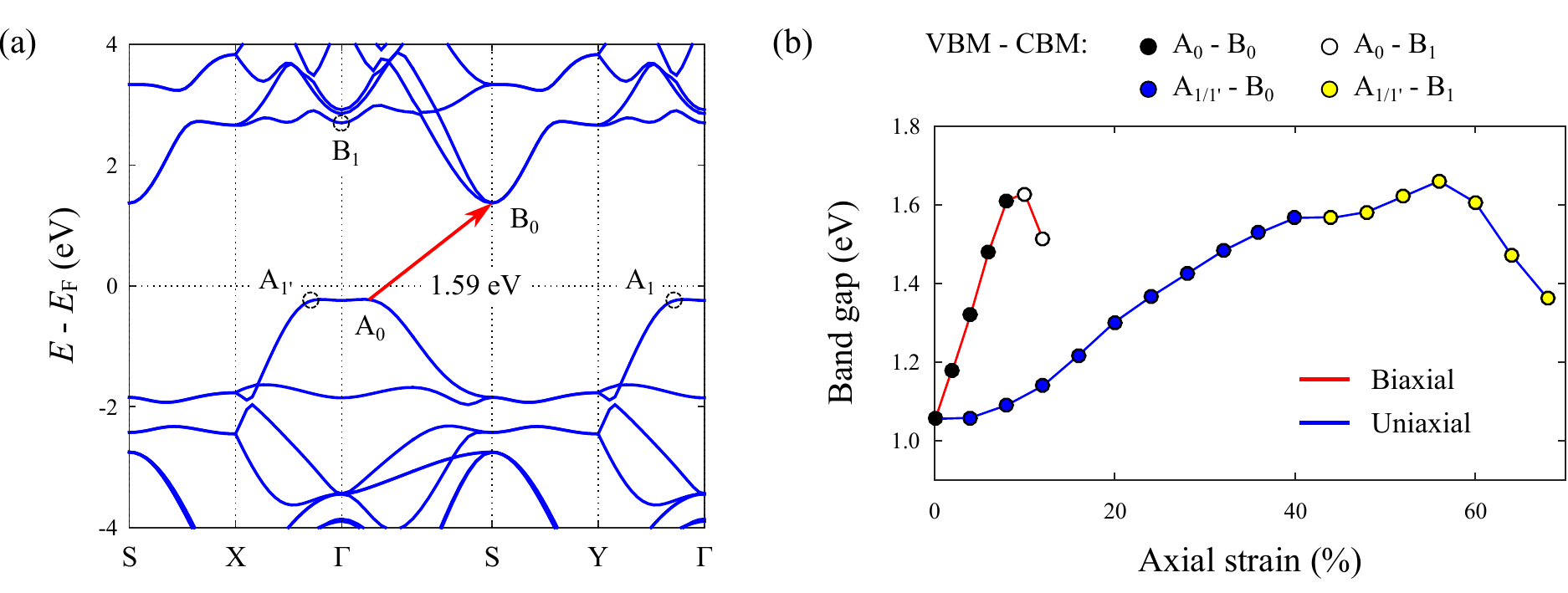}
\caption{(a) Electronic band structure of bilayer SnO in its equilibrium configuration, with reference to the Fermi level \textit{E}\textsubscript{F} (Level of theory: HSE+D2). The near-band-edge states A$_{0}$, A$_{1/1'}$, B$_{0}$ and B$_{1}$ are involved in the VBM-CBM evolution under strain. (b) The variation of the electronic band gap under applied biaxial and uniaxial strain conditions (Level of theory: PBE+D2). The VBM/CBM positions which make up the band gap are denoted accordingly by different colored markers. \label{F5}}
\end{figure*}

\subsection{Strain effect on the electronic properties of bilayer SnO}
Bilayer SnO can be regarded as an ideal prototypical model of tin (II) oxide as it provides a fair representation of interlayer interactions within the system (\textit{d}\textsubscript{bilayer} $\approx$ \textit{d}\textsubscript{bulk} = 2.52 \AA). In this section, the influence of strain on the work function and electronic band structure of ultrathin SnO films is examined by considering its effect in the bilayer. We rely on the PBE+D2 method for sampling the entire strain range of interest, while the use of the more computationally demanding HSE+D2 method is restricted to selected cases for the purposes of validation. Our results indicate that the PBE+D2 approach systematically underestimates the work function and band gap of bilayer SnO, with errors of around 0.45 eV and 0.63 eV, respectively.

The work function \textit{W} is a measure of the minimum energy required to release an electron from a solid to a point in vacuum immediately outside its surface. From the planar-averaged potential plot in Fig. \ref{F4}(a), the equilibrium work function \textit{W}$_0$ of bilayer SnO is determined to be 4.66 eV. As shown in Fig. \ref{F4}(b) (with PBE+D2), for both biaxial and uniaxial deformation modes, the work function generally increases with strain as it approaches their respective regions of critical failure. Refinement with the HSE+D2 method (see Fig. S2(a) in the SM)\cite{SuppMater} reveals that \textit{W} may even exceed 5 eV at high strain levels. Particularly, under uniaxial conditions of below 32\%, we find that its sensitivity to strain is more pronounced than that under relatively higher strain. This phenomenon may even be manifested experimentally in ultrathin SnO films, especially during highly anisotropic mechanical loading, and may be linked to the transition from elastic to plastic deformation of the \textit{h}-SnO intermediate as structural instabilities start to set in.

As shown in Fig. \ref{F5}(a), the indirect band gap in bilayer SnO is 1.59 eV, with the VBM and CBM located at a point along $\Gamma$-S and at the S point, respectively. Note that the S point in reciprocal space is analogous to the M point for the case of equilibrium and biaxial strain conditions wherein tetragonal symmetry is preserved. Changes in the band structure under strain are the result of competition between near-band-edge states A$_{0}$ and A$_{1/1'}$ on the valence band, and B$_{0}$ and B$_{1}$ on the conduction band. The evolution across several indirect band gap types are shown to occur depending on the mode and magnitude of tensile deformation (see Fig. \ref{F5}(b) (with PBE+D2)). Our results are supplemented by calculations with the HSE+D2 method, as shown in Fig. S2(b) in the SM.\cite{SuppMater} Applied uniaxial strain along \textit{x} results in an A$_{1}$-B$_{0}$ band gap, while that along \textit{y} gives rise to an A$_{1'}$-B$_{0}$ band gap, where A$_{1}$ and A$_{1'}$ are located at points along Y-$\Gamma$ and X-$\Gamma$, respectively. For both biaxial and uniaxial deformation modes, the position of the CBM switches from S to $\Gamma$ at very high strains prior to full structural failure. As in the case of the work function, the band gap is also an increasing function of strain within the typical operating range (below the regime of critical failure). The \textit{h}-SnO intermediate, which we report under uniaxial conditions, presents itself as a wide band gap material with high work function, and may bring about interesting opportunities in strain engineering applications. Evidently, manipulating the band gap and work function through strain can allow for greater control of the band alignment and quality of electronic contact at interfaces with ultrathin SnO.

\section{Conclusion}
In this work, via first-principles calculations, we have investigated the effects of tensile strain on the mechanical deformation and electronic properties of ultrathin SnO films. Our results indicate that bulk-to-monolayer transition significantly lowers the Young's and shear moduli of SnO, highlighting the importance of interlayer Sn-Sn bonds in preserving structural integrity. As demonstrated through the case of uniaxial strain, highly anisotropic conditions tend to promote superplasticity, with a critical strain to failure of up to 74\% in the monolayer. Such superplastic behavior is ascribed to the formation of a tri-coordinated intermediate \textit{h}-SnO beyond $\epsilon$ = 14\%, which resembles a partially-recovered orthorhombic phase with relatively large work function and wide indirect band gap. To that end, promoting the stability of \textit{h}-SnO, for example via the selection of suitable substrate materials, surface functionalization and/or environment conditions, may be a necessary and worthwhile direction for future studies. The nature of the indirect band gap of SnO may also be selectively modulated according to the mode and magnitude of tensile deformation as a result of competition between several near-band-edge VBM and CBM states.

The findings reported here provide fundamental insights into the underlying physical mechanism during the deformation process, and strain-driven electronic properties of tin (II) oxide, opening up exciting avenues for the development of SnO-based nanoelectronic devices with new, non-conventional functionalities. This study serves to supplement available literature on atomically thin SnO, while providing a theoretical basis of support for further experimental endeavours which seek to explore SnO as a novel semiconductor material for high-performance TFTs and CMOS devices.

\begin{acknowledgments}
This research article was supported by the Economic Development Board, Singapore and Infineon Technologies Asia Pacific Pte. Ltd. through the Industrial Postgraduate Programme with Nanyang Technological University, Singapore, and the Ministry of Education, Singapore (Academic Research Fund TIER 1-RG174/15). The computational work for this article was partially performed on resources of the National Supercomputing Centre, Singapore (https://www.nscc.sg).
\end{acknowledgments}

\bibliography{bib_manuscript}

\end{document}